\documentclass[prl,twocolumn,reprint,superscriptaddress]{revtex4-2}
\usepackage[utf8]{inputenc}
\usepackage{amsmath}
\usepackage{amssymb}
\usepackage{bm}
\usepackage{graphicx}
\usepackage{graphics}
\usepackage{color}
\usepackage{textcomp}
\usepackage[bookmarks=false]{hyperref}
\usepackage[usenames,dvipsnames]{xcolor}
\usepackage{subfigure}
\usepackage[super]{nth}

\def\ER{ErTe$_3$}
\def\CVS{CsV$_3$Sb$_5$}
\def\AVS{AV$_3$Sb$_5$}
\def\be{\begin{equation}}
\def\ee{\end{equation}}
\def\ba{\begin{eqnarray}}
\def\ea{\end{eqnarray}}

\begin{document}


\title{Thermal transport measurements of the charge density wave transition in CsV$_3$Sb$_5$}


\author{Erik D. Kountz}
\affiliation{Stanford Institute for Materials and Energy Sciences,\\
SLAC National Accelerator Laboratory, 2575 Sand Hill Road, Menlo Park, CA 94025}
\affiliation{Geballe Laboratory for Advanced Materials, Stanford University, Stanford, CA 94305}
\affiliation{Department of Physics, Stanford University, Stanford, CA 94305}

\author{Chaitanya R. Murthy}
\affiliation{Stanford Institute for Materials and Energy Sciences,\\
SLAC National Accelerator Laboratory, 2575 Sand Hill Road, Menlo Park, CA 94025}
\affiliation{Geballe Laboratory for Advanced Materials, Stanford University, Stanford, CA 94305}
\affiliation{Department of Physics, Stanford University, Stanford, CA 94305}

\author{Dong Chen}
\affiliation{Max Planck Institute for Chemical Physics of Solids, 01187 Dresden, Germany}
\affiliation{College of Physics, Qingdao University, Qingdao 266071, China}

\author{Linda Ye}
\affiliation{Geballe Laboratory for Advanced Materials, Stanford University, Stanford, CA 94305}
\affiliation{Department of Applied Physics, Stanford University, Stanford, CA 94305}

\author{Mark Zic}
\affiliation{Geballe Laboratory for Advanced Materials, Stanford University, Stanford, CA 94305}
\affiliation{Department of Physics, Stanford University, Stanford, CA 94305}

\author{Claudia Felser}
\affiliation{Max Planck Institute for Chemical Physics of Solids, 01187 Dresden, Germany}

\author{Ian R. Fisher}
\affiliation{Stanford Institute for Materials and Energy Sciences,\\
SLAC National Accelerator Laboratory, 2575 Sand Hill Road, Menlo Park, CA 94025}
\affiliation{Geballe Laboratory for Advanced Materials, Stanford University, Stanford, CA 94305}
\affiliation{Department of Applied Physics, Stanford University, Stanford, CA 94305}

\author{Steven A. Kivelson}
\affiliation{Stanford Institute for Materials and Energy Sciences,\\
SLAC National Accelerator Laboratory, 2575 Sand Hill Road, Menlo Park, CA 94025}
\affiliation{Geballe Laboratory for Advanced Materials, Stanford University, Stanford, CA 94305}
\affiliation{Department of Physics, Stanford University, Stanford, CA 94305}

\author{Aharon Kapitulnik}
\affiliation{Stanford Institute for Materials and Energy Sciences,\\
SLAC National Accelerator Laboratory, 2575 Sand Hill Road, Menlo Park, CA 94025}
\affiliation{Geballe Laboratory for Advanced Materials, Stanford University, Stanford, CA 94305}
\affiliation{Department of Physics, Stanford University, Stanford, CA 94305}
\affiliation{Department of Applied Physics, Stanford University, Stanford, CA 94305}


\date{\today}

\begin{abstract}
We study thermalization and thermal transport in single crystals of CsV$_3$Sb$_5$  through the CDW transition by directly measuring thermal diffusivity ($D$), thermal conductivity ($\kappa$), resistivity ($\rho$), and specific heat ($c$). Commensurate with previous reports, we observe a sharp, narrow anomaly in specific heat associated with a first order transition that results in a CDW state below $\sim94$ K. While a corresponding sharp  anomaly in thermal diffusivity is also observed, resistivity and  thermal conductivity only exhibit small steps at the transition, where the feature is sharp for resistivity and broader for thermal conductivity.  Scrutinizing the thermal Einstein relation $\kappa=cD$, we find that this relation is satisfied in the entire temperature range, except in a narrow range around the transition. The Wiedemann-Franz law seems to work outside the critical region as well. Below the transition and persisting below the two-phase regime we find strong resemblance between the resistivity anomaly and the specific heat, which may point to a secondary electronic order parameter that emerges continuously below the transition.
 \end{abstract}


\maketitle


\noindent \textbf{\textit{Introduction.}}$-$ Understanding transport in correlated quantum materials is a subject of intense intellectual effort, with newly discovered material systems helping to stimulate fresh and original theoretical work that in particular examine the validity of common assumptions based on kinetic theory. Focusing on thermal transport, this includes  Fourier's law of heat transfer and the Wiedemann-Franz (WF)  law that connects the coefficients of thermal and electrical conductivities of electrons in metals. Where correlations are pronounced and the quasi-particle picture breaks down, both assumptions may fail as was previously demonstrated in studies of thermal transport beyond the Mott-Ioffe-Regel (MIR) limit in cuprates \cite{Zhang2017,Zhang2019,Zhang2019b}. There, phonons were shown to play a key role, which was strikingly  evident from parallel studies of diffusivity bounds in similar complex insulators \cite{Behnia2019}. 

Another fertile ground to examine non-quasiparticle transport is in the vicinity of a charge density wave (CDW), which is driven by an interplay of both strong electron-electron and electron-phonon interactions. For example, thermal diffusivity, resistivity, and specific heat measurements on the CDW material \ER~ exhibit a sharp decrease in thermal conductivity both parallel and perpendicular to the primary CDW at the CDW transition temperature, while the resistivity changes more gradually, implying a strong breakdown of the WF law in the critical regime of the CDW transition \cite{Kountz2021}. Furthermore, assuming Fisher-Langer (FL) theory \cite{Fisher1968} applies to the continuous CDW transition in \ER, large anomalies observed in the temperature derivative of the resistivity stands in sharp contrast to the small anomaly observed in heat capacity measurements \cite{SaintPaul2017}. These results are suggestive of a phenomenological recount of a strongly coupled electron-phonon critical ‘soup’   \cite{Kountz2021}. 

A particularly exciting new material system, where a CDW state appears as a forceful effect, is the class of quasi-two-dimensional kagom\'e metals \AVS, which exhibit charge order transitions at $\sim$80 K, 103 K and 94 K for A = K, Rb and Cs respectively \cite{Ortiz2020}.  Focusing on \CVS, the CDW transition is associated with a reconstruction of the Fermi surface pockets linked to the vanadium orbitals and the kagom\'e lattice framework \cite{Ortiz2021}. Nuclear magnetic resonance (NMR) studies on the different vanadium sites are consistent with orbital ordering at $T_{CDW}\sim$94 K induced by a first order structural transition, accompanied by an electronic charge density wave (CDW) that appears to grow gradually below $T_{CDW}$, with possible intermediate subtle stacking transitions perpendicular to the Kagom\'e planes \cite{Song2022}. With superconductivity appearing at lower temperatures ($\sim 4$ K), \CVS~ is a prime example of a system exhibiting ``intertwined order’’ \cite{Fradkin2015,Fernandes2019}, where multiple phases emerge out of a primary phase, starting from the first order transition at $\sim$94 K.  Understanding the consequences of this unique electron-phonon landscape on thermal transport is the primary objective of the present study.

In this letter we examine the dynamics of the CDW phase transition by means of independent measurements of the specific heat, thermal conductivity, thermal diffusivity and electrical resistivity. Our primary result is that thermal transport in a critical regime near $T_{CDW}$ shows anomalous behavior inconsistent with quasiparticle transport.  In addition, we also observe:  {\it i}) The specific heat shows a strong singularity at $T_{CDW}$, although no hysteresis could be detected;  {\it ii}) Despite it being a first order transition, a  finite drop in the resistivity at $T_{CDW}$ yields a strong peak in its temperature derivative, which seems to correspond to the specific heat anomaly similar to continuous CDW transition  materials \cite{Fisher1968}. {\it iii}) The thermal conductivity is almost temperature independent above the transition, but starts to increase below the transition; {\it iv}) Thermal diffusivity through the CDW transition shows a sharp decrease, which cannot be fully explained by the ratio $D=\kappa/c$; {\it v}) except for the region of $\sim 2$ K around the CDW transition, the relation $\kappa=cD$ is satisfied quantitatively and applying the WF law to the electrical resistivity, the difference thermal conductivity gives a reasonable result for the phonon contribution to the heat transport.
\bigskip

\begin{figure}[ht]
\centering
\includegraphics[width=0.86\columnwidth]{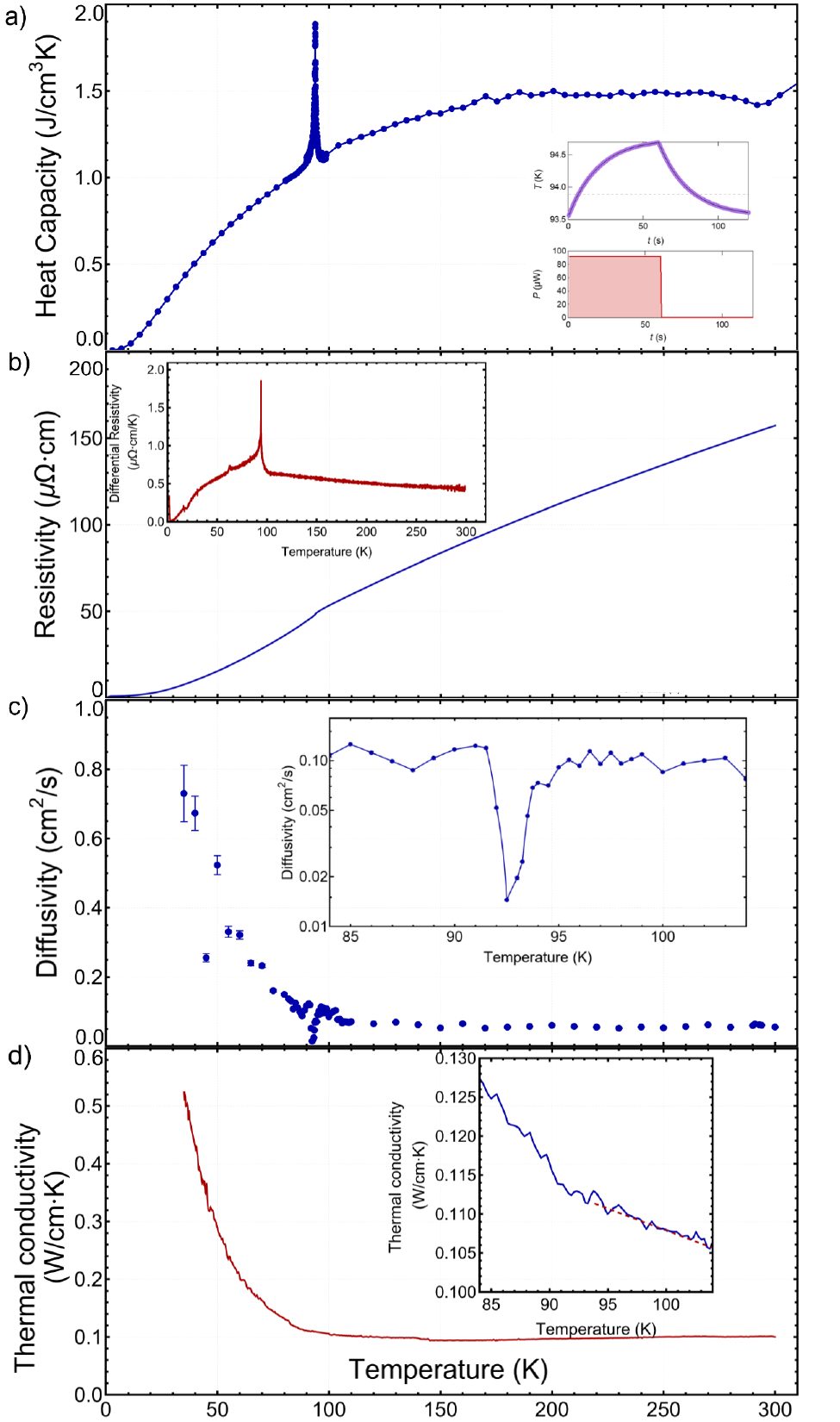}
\caption{\textbf{Electrical and thermal measurements on CsV$_3$Sb$_5$ crystals:} (a) Rescaled specific heat vs. temperature. Note the only anomaly in specific heat is at $T_{CDW}$. Inset: temperature as a function of time through $T_{CDW}$ with heater on (before 60~s) and with heater off (after 60~s). Note no plateau of temperature indicating no latent heat. (b) Rescaled a-b plane resistivity  vs. temperature. Inset: differential resistivity vs. temperature. Note large anomalies at $2.5$~K (superconductivity) and $94$~K (CDW). There are smaller anomalies at $62$~K possibly indicating a transition to a $2\times2\times4$ supercell and at $18$~K.  (c) Thermal diffusivity measured using photothermal microscope showing CDW transition at $T_{CDW}\approx 94~$K. Scatter of data primarily associated with one-pixel control of measurement position and relative distance between heating and probing laser spots. Note the sharp drop of diffusivity at the CDW transition and the strong increase below the transition. (d) Direct measurement of total thermal conductivity $\kappa$ vs. temperature rescaled to match known room temperature values. See \cite{SuppMat} for details. Inset: thermal conductivity in the vicinity of $T_{CDW}$. }
\label{all}
\end{figure}

\noindent \textbf{\textit{Results.}}$-$ CsV$_3$Sb$_5$ samples were grown at MPI Dresden following the procedure described in \cite{Huang2022}, \cite{Ortiz2020}, and in related publications.  Here a self-flux method was used, ensuring melt purity, and producing large crystals with a high degree of structural order. For the thermal diffusivity measurements we used a photothermal microscope \cite{Zhang2017}. Fig.~\ref{all} shows specific heat, resistivity, thermal diffusivity and thermal conductivity measurements on same-batch CsV$_3$Sb$_5$ crystals. Technical details of these measurements are described in the Supplementary Material (SM)\cite{SuppMat}. 

Figure~\ref{all}(a) shows specific heat of a same-batch CsV$_3$Sb$_5$ crystal (see SM \cite{SuppMat} for determination of geometrical factors), closely matching previous work \cite{Ortiz2020,Li2021} featuring a strong anomaly at $T_{CDW}$ with the magnitude of the anomaly over 66\%. Fitting the specific heat to a Debye model, it shows an increase in Debye temperature from $\theta_D=160$~K at $10$~K to $\theta_D=260$~K at $80$~K and above through the CDW transition, and saturates at the high temperature Dulong-Petit value. The specific heat at $T_{CDW}$ almost doubles its value, showing a sharp anomaly that rises to $\Delta c_p\approx 0.8~$J/cm$^3\cdot$K within $\sim$3 K of the transition. However, despite its first-order nature, no hysteresis or latent heat (down to $6.6\times 10^{-4}$~J$/$cm$^3$) was detected through the CDW transition (see Fig.~\ref{all}(a) inset).

CsV$_3$Sb$_5$ resistivity was measured previously. However, due to the strong anisotropy of this layered structure, it is difficult to quantitatively determine the resistivity from resistance measurements. Indeed, scale of resistivity reported in the literature varies between $31~\mu\Omega\cdot$cm and $300~\mu\Omega\cdot$cm \cite{Ortiz2020,PRBWang2021,Xiang2021,Li2021,Huang2022,Chen2022,Zhou2022,Uykur2021,Zhou2021,Chen2021}, although all resistivity measurements can be scaled to show the same temperature dependence, including a small ``jump'' at $T_{CDW}$.  Fig.~\ref{all}(b) shows resistivity measurements on our crystals exhibiting $160~\mu\Omega\cdot$cm at $300$~K and a RRR between 150 and 220, resulting in very low residual resistance. We can also use the WF law and our thermal diffusivity and conductivity measurements to put a lower bound of $\sim40\%$ for the heat conducted by electrons at room temperature, while the remainder, i.e. phononic, component of thermal conductivity $\sim 0.06$~W$/$cm$\cdot$K is similar to other layered chalcogenide materials \cite{Kountz2021}. Using WF to extract electronic thermal conductivity below $T_{CDW}$ yields a sharp upturn, before it turns down at low temperatures. Finally, the small resistivity jump at $T_{CDW}$ yields a very sharp singularity in the derivative $\partial\rho/\partial T$ shown in Fig.~\ref{all}(b) inset. A weaker anomaly seen at $\sim62$ K may correspond to a previously observed transition to a $2\times2\times4$ supercell \cite{Ortiz2021}, while an observed anomaly at $\sim18$ K was not previously reported. 

\begin{figure}
\centering
\includegraphics[width=1.0\columnwidth]{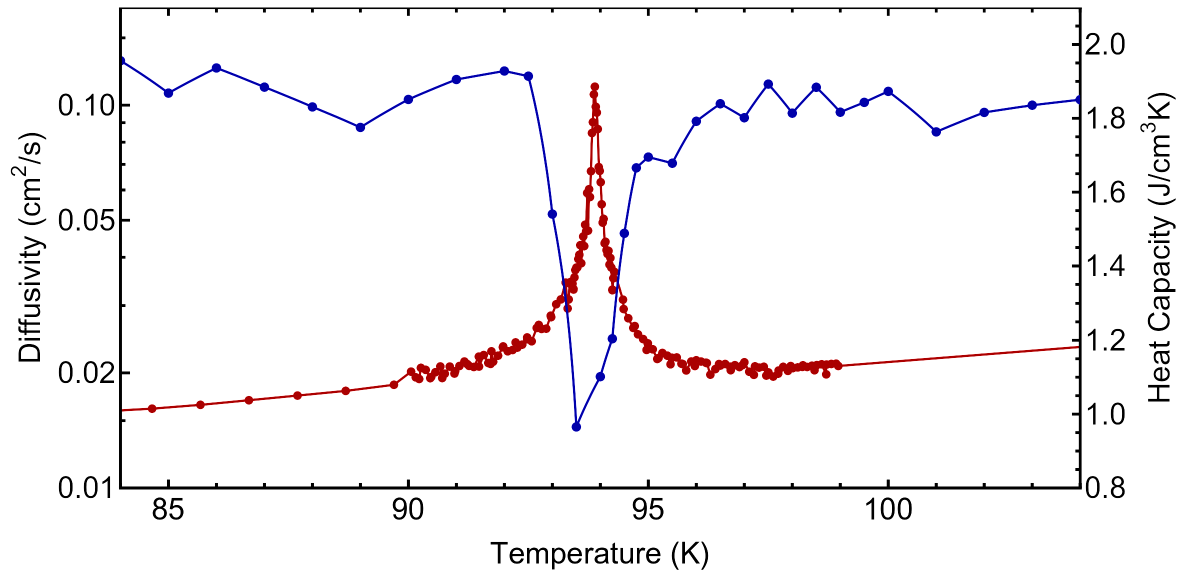}
\caption{CsV$_3$Sb$_5$ thermal diffusivity (left axis) measured using photothermal microscope near $T_{CDW}\approx 94~$K (blue points). Diffusivity values shifted up by $1$~K from different same-batch crystals in different apparatuses to match $T_{CDW}$ with specific heat. Scatter of data primarily associated with one-pixel control of measurement position and relative distance between heating and probing laser spots. Additional 5\% systematic uncertainty, see SM \cite{SuppMat} not included. Note the increase in diffusivity below $T_{CDW}$ and the very sharp drop of diffusivity by over an order of magnitude at the CDW transition. Red curve shows specific heat (right axis). Note that while the diffusivity decreases by $90$\%, specific heat only increases by $66$\%.}
\label{cdiff}
\end{figure}

Thermal diffusivity, shown in Fig.~\ref{all}(c),  was directly measured using a photothermal microscope \cite{Zhang2017} from $30$~K to $300$~K, showing three different regions: below $T_{CDW}$, around $T_{CDW}$, and above $T_{CDW}$. Above the CDW transition temperature, the diffusivity is approximately constant with a slight increase at lower temperatures. Lowering the temperature towards the transition, the thermal diffusivity sharply decreases by at least a factor of $\times20$ as it is limited by the resolution of the measurement. This anomaly is followed by a recovery to roughly the value just above $T_{CDW}$, with a trend of increasing thermal diffusivity below the transition. Fig.~\ref{cdiff} depicts an expanded region of the CDW transition, showing that within the larger uncertainty in diffusivity data, the specific heat and diffusivity anomalies are roughly similar in width, about $\lesssim3$ K around $T_{CDW}$, similar to previously reported NMR measurements \cite{Song2022}. Below $T_{CDW}$ a sharp increase in thermal diffusivity together with a decrease in electrical resistivity is consistent with decrease in electron-phonon scattering.

Thermal conductivity in the $a$-$b$ plane was also measured as shown in Fig.~\ref{all}(d) with the aim to compare this direct measurement to the measurements of specific heat and diffusivity. Here a known heat current was introduced to the sample and the measured temperature gradient across the sample was measured, showing values between $0.05$~K and $0.8$~K. Additional details of thermal conductivity measurements are described in the Supplementary Material (SM): \cite{SuppMat}. The most striking observation here is that the thermal conductivity does not show any singularity at the CDW transition, similar to previously reported results in \cite{Zhou2022}. The inset in Fig.~\ref{all}(d) further indicates a tendency to an increase below $T_{CDW}$, again similar to \cite{Zhou2022}.  Assuming as in Eqn.~\ref{eqn1} a global thermal ``Einstein relation'' for the thermal conductivity $\kappa=cD$, it is surprising that vestiges of the first order transition singularity at $T_{CDW}$ are absent in $\kappa$.

\begin{figure}[h]
\centering
\includegraphics[width=1.0\columnwidth]{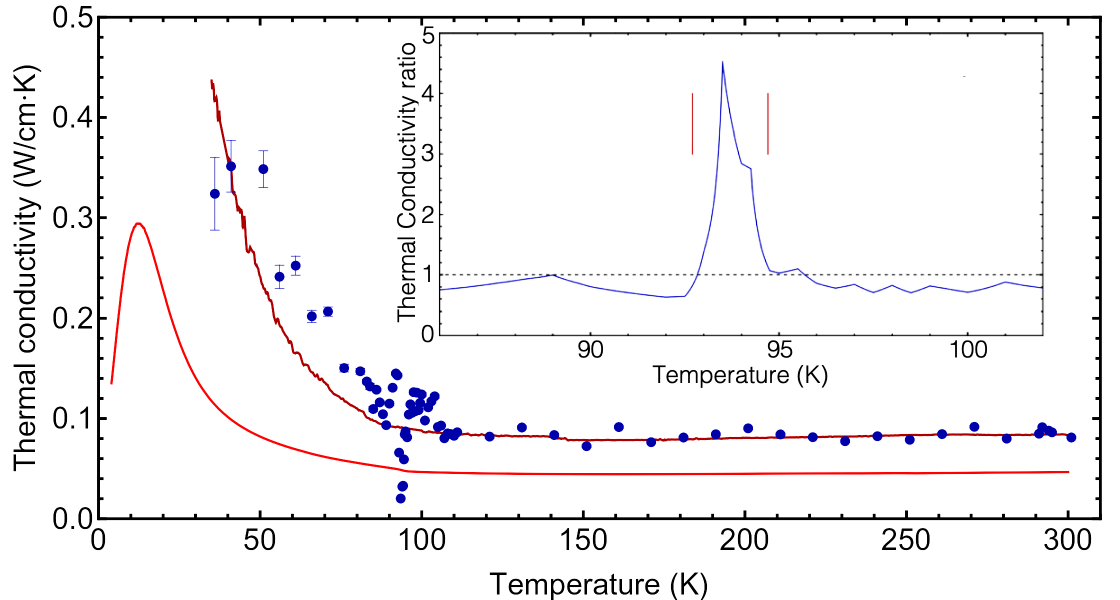}
\caption{Direct measurement of total thermal conductivity $\kappa$, of CsV$_3$Sb$_5$ (dark red and dark orange), electronic component $\kappa_{el}$ computed from $\rho$ assuming WF law (red), and calculated conductivity with $1$~K shift in diffusivity to match $T_{CDW}$ with specific heat $\kappa=Dc$ (dark blue). Inset shows the ratio: $R_q=\kappa/cD$ as a function of temperature indicating a strong deviation from the simple hydrodynamic regime ($R_q=1$) in the vicinity of the CDW transition (marked with two vertical lines). Note the otherwise good agreement between the directly measured and calculated conductivities.}
\label{kappa}
\end{figure}
Comparing the direct measurement of thermal conductivity with the aforementioned thermal Einstein relation, Fig.~\ref{kappa} shows an overall excellent correspondence of the two approaches above the critical region, a strong deviation in the critical region and a small deviation below the CDW transition.   An instructive way to demonstrate this observation is shown in the inset of Fig.~\ref{kappa} where we plot the ratio: $R_q\equiv \kappa/cD$ using the experimentally measured quantities. Where the kinetic approach prevails, we expect $R_q=1$, which seems to hold above $T_{CDW}$. We find that $R_q\gg 1$ in the $\sim3$ K around $T_{CDW}$, which is the two phase regime as was also found in  the NMR experiments of Song {\it et al.}, \cite{Song2022}.  We also include in Fig.~\ref{kappa} the electronic thermal conductivity calculated from the measured resistivity and assuming the WF law, which shows similar trend outside the CDW transition region, with the difference from the measured thermal conductivity indicating a reasonable phonon contribution, at least above $T_{CDW}$ \cite{Kountz2021}. Below $T_{CDW}$ the difference $\Delta\kappa=\kappa-\kappa_e$ shows increase as well, presumably due to increase in both phonon and electron thermal conductivities due to decrease in electron-phonon scattering. It is thus reasonable to assume that $\Delta\kappa\equiv\kappa_{ph}$ outside the critical regime.

\noindent  \textbf{\textit{Discussion.}}$-$ Heat conduction in solids is mediated by the thermal motion of quasiparticles and  various elementary excitations, primarily phonons, which also serve to locally thermalize the solid. Where both electrons and phonons are well defined, the thermal conductivity is the sum of both electrons and phonons thermal conductivities $\kappa=\kappa_e+\kappa_{ph}$, such that a temperature gradient applied to the sample results in a heat flux density 
\begin{equation}
\vec{j}_q=-\kappa\nabla T,
\label{fourier}
\end{equation}
transferring entropy from the hotter to the colder end of the sample. This so-called Fourier's law considers heat as a ``fluid'' and together with the first law of thermodynamics lead to the heat equation.  Applying energy conservation, the local 
molar energy density $u(\vec{r},t)$ will satisfy a continuity equation
\begin{equation}
\frac{\partial u}{\partial  t}+\nabla\cdot\vec{j}_u=0
\label{conti}
\end{equation}
Since most experiments are done at constant pressure conditions, it is advantageous to consider the first law in the form $dh=Tds+vdP+\sum_i\mu_idn_i$, where $h(s,P,\{n_i\})$ is the molar enthalpy of the system, $s$ is the molar entropy, $v$ is the molar volume, $P$ is the pressure, $\mu_i$ is the chemical potential of the $i^{th}$ component (if any) and $n_i$ is the respective density. For a one-component system such as in our present experiment and under constant pressure and density conditions, it is advantageous to consider the continuity equation for the enthalpy. This can be shown to yield the heat equation (written in components convention):
\begin{equation}
c_p\left(\frac{\partial T}{\partial t}\right)_P-\kappa_{ij}\partial_i\partial_j T-(\partial_i\kappa_{ij})\partial_j T=0
\label{heatdiff}
\end{equation}
For most cases the thermal conductivity is uniform in any particular direction in space and thus the second gradient term vanishes yielding a simple diffusion equation with  $D_{ij}=\kappa_{ij}/c_p$. If the electrons and phonons are in equilibrium and well defined quasiparticles, the kinetic approach (e.g. for the isotropic case) implies:
\begin{equation}
\kappa=c_pD=\kappa_e+\kappa_{ph}=c_eD_e+c_{ph}D_{ph}
\label{eqn1}
\end{equation}
 where $c_e$ and $c_{ph}$ are the electronic and phononic specific heats and $D_e$, $D_{ph}$ are the corresponding thermal diffusivities. Furthermore, when transport is dominated by weakly interacting 
 quasi-elastic scattering processes, $\kappa_{e}$ is related to electrical conductivity by the Wiedemann-Franz law, i.e. $\kappa_{e}/\sigma=L_0T$, where  $L_0=\pi^2k_B^2/3 e^2~\approx 2.44\times 10^{-8}~$W$\Omega$K$^{-2}$ is a universal constant. Observing this ratio indicates ``standard'' transport in a given electronic system, while significant violations of the WF law may indicate a breakdown of the quasiparticle description (see e.g. \cite{Mahajan2013}). 

The data presented above clearly shows that while the kinetic approach, which follows Eq.~\ref{eqn1} holds in most of the temperature regime, it strongly fails in a narrow temperature range around the CDW anomaly. We noted earlier that $R_q=1$ holds except for a $\sim3$ K range around $T_{CDW}$. In fact, this is the exact temperature range where a two-phase region is observed in the NMR experiments of Song {\it et al.}, \cite{Song2022}.  
Since the thermal conductivity does not show any significant anomaly at $T_{CDW}$ (similar to the resistivity), we would a-priori expect that the anomaly in the specific heat will be compensated by the anomaly in the thermal diffusivity. The fact that $R_q$ shows a strong singularity at the transition indicates the breakdown of simple hydrodynamic heat diffusion. This can happen if e.g. the nonlinear term in Eqn.~\ref{heatdiff} becomes significant in the two-phase regime where  the system exhibits randomly distributed puddles of the two phases, thus $\kappa_{ij}$ varies in space, causing the $\partial_i\kappa_{ij}$ term to be finite.  In addition to the spatial inhomogeneity in the two-phase regime,  the latent heat of the transition \cite{Agne2019} will create local temperature variations, which will be time dependent as per the first term in Eqn.~\ref{heatdiff}. Thus, at a given time scale the term  $(\partial_i\kappa_{ij})\partial_jT$ may dominate over simple diffusion, yielding a strong anomaly in the measured diffusivity. For a slow measurement of the thermal conductivity the electrons and phonons are in equilibrium and $\kappa$ may not acquire any singularity through the transition. However, the diffusivity is a dynamical property that depends on the time scale of the measurement, the associated relative fraction of the two phases through the transition as well as their spatial distribution. 

Focusing on the experimental results, both the specific heat (Fig.~\ref{all}a) and the thermal conductivity  (Fig.~\ref{all}d) were measured quasi-statically as is evident from the lack of hysteresis in these two measurements. On the other hand the thermal diffusivity was measured at a finite frequency (1 to 5 kHz), exhibiting a very strong dip at the transition and accompanied by a large scatter of the diffusivity value in different runs, presumably due to difference in local distribution of the two phases.  Below the transition the system is anisotropic and whether $R_q=1$ may depend strongly on the nature of the CDW state. Indeed, close examination of directly measured $\kappa$ vs. the combination $cD$ below $T_{CDW}$ may indicate the latter exhibiting a slightly larger value and $R_q\lesssim1$. However, as seen in Fig.~\ref{kappa}, the steep increase of the thermal conductivity below the transition prevents an accurate determination of this value. In particular we note the large variation in resistivity among measured samples, which if WF is used implies similar variation in the electronic part of the thermal conductivity.

\begin{figure}[ht]
\centering
\includegraphics[width=1.0\columnwidth]{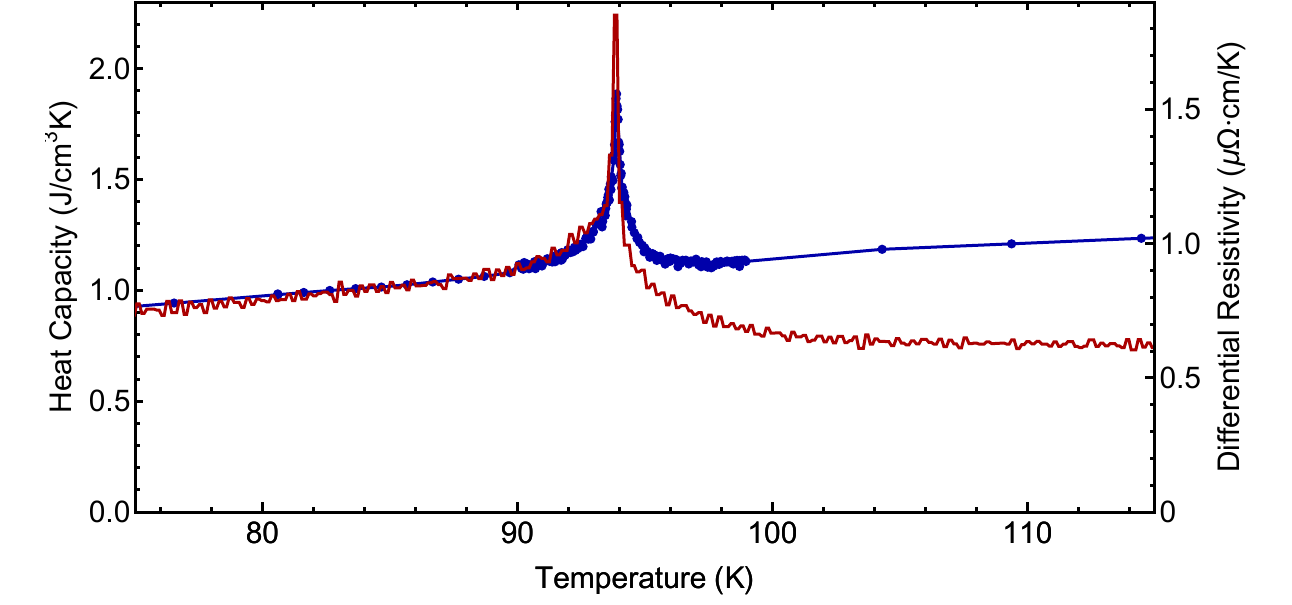}
\caption{Differential resistivity $d\rho/dT$ vs. temperature and specific heat vs. temperature near $T_{CDW}$. Note the strong correspondence below $T_{CDW}$ but not above. }
\label{FL}
\end{figure}
Turning to the anomaly of the electronic response through the transition, it is instructive to plot the derivative of the resistivity, which exhibits a sharp drop at the transition. We notice that while above the transition the correspondence of the two anomalies is poor, below the transition it is remarkably similar, persisting well below the narrow two-phase regime.  For a continuous transition Fisher-Langer (FL) theory  \cite{Fisher1968} states that the dominant contribution to the scattering near the phase transition of a metal is due to the short-range order-parameter fluctuations, which in turn cause the temperature derivative of the resistivity through the transition to simulate the specific heat anomaly. However, here we observe a first order transition, for which the FL theory is not applicable.  A possible explanation to the correspondence of the anomalies in $d\rho/dT$ and $c_p$ below the transition is that a corresponding continuous order parameter emerges, which is triggered by the first order distortion transition at $T_{CDW}$. This idea is further supported by NMR studies, which suggest that the observed transition originates from orbital ordering at $T_{CDW}\sim94$~K induced by a first order structural transition, accompanied by electronic charge density wave (CDW) that appears to grow gradually below $T_{CDW}$ \cite{Song2022}. In that case we may expect two different effects that will affect the resistivity through the transition. While the first order structural transition may primarily induce the finite step in the resistivity as a result of a sharp change in the density of states, the accompanied continuous formation of CDW below the transition may be responsible for the enhanced scattering, conforming to the FL theory. 

In conclusion by comparing direct measurements of thermal conductivity, specific heat, thermal diffusivity and electrical resistivity we could test the validity of the kinetic approximation, where heat is considered a hydrodynamic diffusive fluid, as well as the Wiedemann-Franz law for the electronic part of the thermal conduction. We find that both hallmarks of standard transport in metals hold except in the very narrow regime of the first-order transition, where a two-phase regime appears.

\section{Acknowledgments}
\begin{acknowledgments}
This work was supported by the Department of Energy, Office of Basic Energy Sciences, under contract no. DE-AC02-76SF00515. 
\end{acknowledgments}

\bibliography{thermal}

\newpage

\onecolumngrid
\newpage
\setcounter{section}{0}
\setcounter{figure}{0}
\renewcommand{\thefigure}{S\arabic{figure}}
\renewcommand{\theequation}{S.\arabic{equation}}
\renewcommand{\thetable}{S\arabic{table}}
\renewcommand{\thesection}{S\arabic{section}}

\renewcommand{\thefootnote}{\fnsymbol{footnote}}

\begin{center}
\textbf{ SUPPLEMENTARY INFORMATION}

\vspace{3em}
\textbf{Thermal transport measurements of the charge density wave transition in CsV$_3$Sb$_5$}\\

\fontsize{9}{12}\selectfont

\vspace{3em}
Erik D. Kountz$^{1,2,3}$, Chaitanya R. Murthy$^{1,2,3}$, Dong Chen$^{4,5}$, Linda Ye$^{2,6}$, Mark Zic$^{2,3}$, Claudia Felser$^4$, Ian R. Fisher$^{1,2,6}$, Steven A. Kivelson$^{1,2,3}$, Aharon Kapitulnik$^{1,2,3,6}$

$^1${\it Stanford Institute for Materials and Energy Sciences,
SLAC National Accelerator Laboratory, 2575 Sand Hill Road, Menlo Park, CA 94025}\\
$^2${\it Geballe Laboratory for Advanced Materials, Stanford University, Stanford, CA 94305}\\
$^3${\it Department of Physics, Stanford University, Stanford, CA 94305}\\
$^4${\it Max Planck Institute for Chemical Physics of Solids, 01187 Dresden, Germany}\\
$^5${\it College of Physics, Qingdao University, Qingdao 266071, China}\\
$^6${\it Department of Applied Physics, Stanford University, Stanford, CA 94305}
\bigskip
\end{center}

\section*{ MATERIALS, METHODS AND ADDITIONAL INFORMATION}


\subsection*{Single Crystal Growth}

CsV$_3$Sb$_5$ single crystals were prepared by self-flux which ensures purity of the melt and produces large crystals with a high degree of structural order. The measured crystal sizes were on the order of $1\times2\times0.01$ mm$^3$. All measured crystals came from same batch growths. Crystals were grown with slight excess of Cs \cite{Ortiz2020}.

\subsection*{Specific heat measurements}

The heat capacity of the single-crystal samples was measured using a relaxation time technique in a Quantum Design Physical Property Measurement System (PPMS). A crystal with a mass of approximately $400~\mu$g and dimensions approximately $1\times3\times0.011$ mm$^3$ with flat surfaces was selected for good thermal contact with the sample platform. Data were taken in zero applied field from $2.9$ to $300~$K. The dc temperature increase was $2\%$ at all temperatures with additional measurements at $1\%$ and $0.5\%$ temperature increases near $T_{CDW}$. Specific heat was converted from J/K to J/cm$^3 \cdot $K by measuring the sample thickness and sample area with a microscope to calculate area and using density from the unit cell from single-crystal x-ray diffraction (SCXRD) of $6.102$~g$/$cm$^3$\cite{Ortiz2019}.

\subsection*{Resistivity measurements}

Because CsV$_3$Sb$_5$ has hexagonal symmetry both in and out of the CDW state, axis of measuring the resistivity, diffusivity, and conductivity was not aligned with XRD. However, the axis was the constant for each measurement. Multiple measurements of resistance versus temperature were performed with a traditional dipping probe measurement and with a PPMS. The resistivity in the $a-b$ plane was measured on thin rectangular crystals which had been cut with a scalpel before contacting with silver paint. The largest source of error in such resistivity measurements is the magnitude of the resistivity resulting from the uncertanty in sample dimensions particularly sample thickness. Multiple measurements of resistivity were measured and the smallest resistivity measurement was used as the reported resistivity in the $a-b$ plane. Other resistivities were rescaled to this reference resistivity presented in this letter. This system of reporting the lowest resistivity was done because in many samples the current contact do not make contacts to the entire depth of the sample but instead only the topmost layers. In all cases, the resistivity data measured was higher than previous measurements from MPI \cite{Huang2022}. If the data was not rescaled the resulting electrical thermal conductivity would be larger and the phononic thermal conductivity would be smaller than presented. Differential resistivity was calculated by taking the numerical derivative of the resistivity data.

\subsection*{Thermal conductivity measurements}

Direct measurements of total thermal conductivity versus temperature were performed using a dc heating measurement approach. A smooth flat crystal of CsV$_3$Sb$_5$ was cut to a rectangular shape. The bottom of the crystal was vertically mounted to a copper block with silver paste right next to an Omega $100~$Ohm Pt thermometer to measure the sample temperature. The copper block was then attached to the copper cryostat coldfinger with GE varnish and cigarette paper to electrically isolate the sample. On the top of the CsV$_3$Sb$_5$ crystal, a $350~$Ohm strain gauge was attached to act as a heater. Current leads used $40~$AWG copper wire. On the sides of the CsV$_3$Sb$_5$ sample $2~$mil type T thermocouples were attached with GE varnish. The ends of the thermocouples were soldered to pins which were connected to the outside of the cryostat by manganin wires to reduce thermal heat flow. The soldered connections were thermally anchored with GE varnish and cigarette paper on the coldfinger close to the Pt thermometer to minimize temperature gradients and thermoelectric voltages generated at the junctions. All wires were chosen to be over $3~$cm long to minimise any thermal leakage. Thermoclectric voltages were measured with Keithley 181 and 182 nanovoltmeters.

The sample was cooled down to 30~K with liquid helium or 80~K with liquid nitrogen the setup slowly warmed or cooled. Measurements of thermoelectric voltages, temperature, heating current, etc. were taken every $\sim~1.5~$s giving a heating/cooling rate of $1~$K$/5~$min (200 measurements per degree), $1~$K$/10~$min (400 measurements per degree), or $1~$K$/15~$min (600 measurements per degree). The heating current was supplied with a Keithley 220 current supply and voltage across the heater measured with a Keithley 197A multimeter. After every measurement, using the measured voltage across the heater, the current supplied was changed so that a constant amount of power was applied to the crystal. 

Every 50-200 measurements, the heating current is switched between $1~\mu$W and the chosen heating power: $0.1~$mW, $0.2~$mW, $0.5~$mW, or $1~$mW. A ``no heating power'' of $1~\mu$W was chosen so that the resistance of the resistor would always be known. Heating from this small current was not discernible from cases where no $0~\mu$W of heating was applied. After dropping 6-16 data points around when the heating current was switched to avoid transients, the data for each heating power measurement was averaged and interpolated for all temperatures to find the temperature difference between each thermocouple and the base temperature. Then, the temperature difference between the two thermocouples for heating current on and off was seperately calculated by subtracting the previously calculated temperature differences. Finally, the temperature gradient from the heating current was calculated by subtracting the `heater on' temperature difference from the `heater off' temperature difference. This second subtraction is used to eliminate any residual thermoelectric voltages across any junctions elsewhere in the cryostat to get the true temperature gradient. This temperature gradient is then inverted to find the thermal conductance. The thermal conductivity is calculated by multiplying the thermal conductance by the sample dimensions and a global temperature independent rescaling factor so that the room temperature conductivity matches that of the conductivity calculated from diffusivity.

Resulting temperature gradients were approximately $0.1~$K to $0.05~$K depending on the applied heating current.

\subsection*{Thermal diffusivity measurements}

Thermal diffusivity measurements were performed using a photo-thermal microscope working in reflection mode first introduced for the study of single crystal cuprate superconductors by Fanton {\it et al.} \cite{Fanton1989}.  A complementary comprehensive study of the technique was given by Hua {\it et al.} \cite{Hua2015}. The specifics of the apparatus used in the present study is first described in \cite{Zhang2017}. Using this apparatus, the thermal diffusivity is obtained directly, without the need to measure the thermal conductivity and specific heat separately. An advantage of this apparatus is the ability to measure the full any in-plane anisotropy of the thermal diffusivity by orienting the pair of heating and probing laser spots at any arbitrary orientation with respect to the crystal axes. The mobility in the optics is further used for diagnostics of spatial uniformity of the thermal diffusivity. 

\subsubsection*{Principles of the Photothermal Apparatus}

\begin{figure}[ht]
	\centering
	\includegraphics[width=1.0\columnwidth]{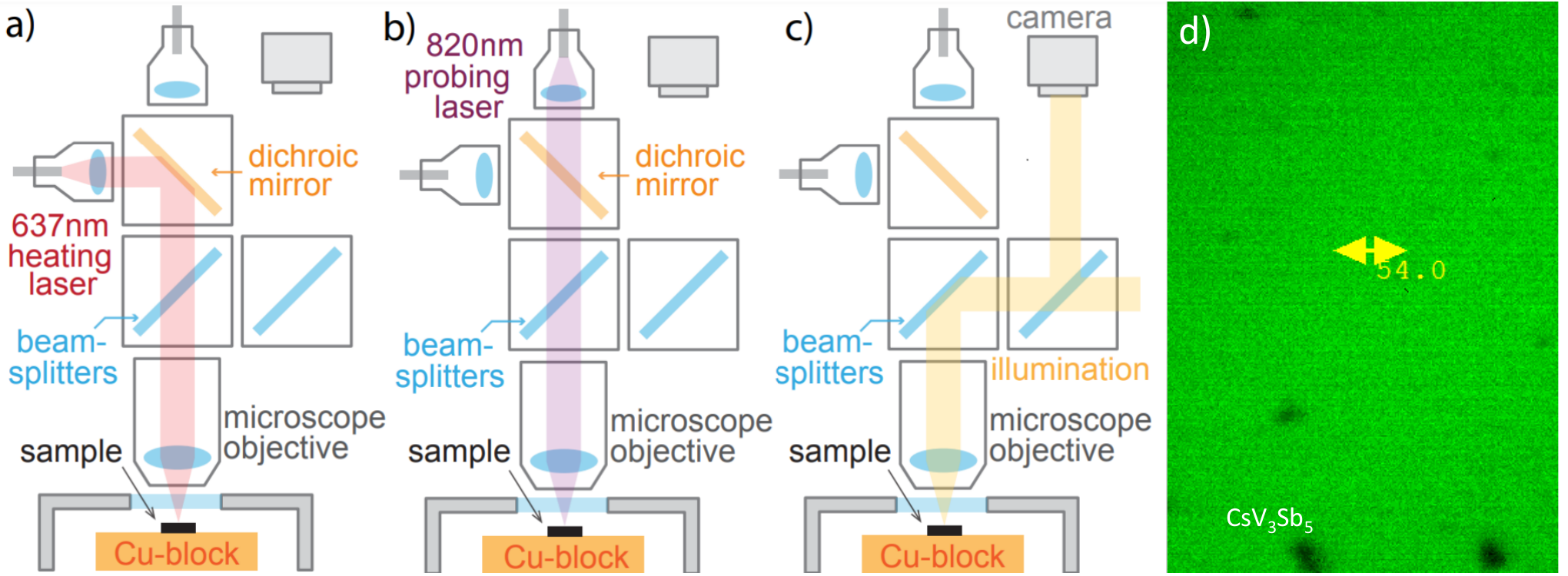}
	\caption{\label{setup}(color online) The schematic shows the optical paths of the setup. (a) Path of the heating laser (637~nm and 642~nm) (b) Path of the probing laser (820~nm). The reflected light traverses the same path before gathered by a photodetector. (c) Path of illumination light source and camera vision. (d) Location of focused laser spots at typical measurement separation distance (distance given is 54 pixels or 9.99~$\mu$m. The false color screenshot is taken on a CsV$_3$Sb$_5$ sample surface for visual interest, on an atomically flat surface with a few spots in the bottom of the image can be seen. The surface of crystals measured in the main text are mostly featureless.}
\end{figure}

For the high resolution thermal diffusivity measurements we use a home-built photothermal microscope.  The microscope views the sample through a sapphire optical window in a cryostat, with the sample mounted to a cold finger just under the window.  A schematic is shown in Fig.~\ref{setup}. A heating laser at $637~$nm or $642~$nm and a probing laser at $820~$nm are focused onto the sample surface by the microscope objective.  The focused spots have Gaussian size of approximately $1~\mu$m and $2~\mu$m, respectively, due to the diffraction limit of different wavelengths, and can be moved independently over the sample surface.  A camera allows us to observe the sample surface nearby, align the spots in a particular orientation with respect to the crystal, and determine the distance between the spots.  

The output power of the heating laser is modulated with a sinusoidal profile $P(t)=P_0 [1 + \sin(\omega_0 t)]$. The modulation frequency $\omega_0/2\pi$ has a typical range of $5.5~$kHz - $25.5~$kHz, much slower than the microscopic equilibration on the order of picoseconds. This means that the parameters extracted are all within the in the DC limit of linear response, and that the dependency on the modulation frequency can be neglected. The probing laser is aimed at a spot a small distance (typically $9-15~\mu$m) away from the heating laser. The reflected light from the probing laser is diverted by an optical circulator and fed into a photodetector. The AC component of the photodetector signal is then fed to a lock-in amplifier referenced to the laser modulation and the amplitude and phase are measured. Modulation of the heating laser at a frequency $\omega$ implies that at the detector $\delta R \propto (dR/dT)\delta T$. The DC component is used as a gauge to make sure the lasers are focused. Since the electronic thermalization time is many orders of magnitude faster than the heat-modulation frequency, the heat wave that emerges from the heating spot is carried by both electrons and phonons, which are in thermal equilibrium with each other, propagating in the radial direction. Given the sample's total specific heat capacity, the phase shift that is measured at the probing spot reflects the thermal diffusivity of the sample. We further emphasize that the local optical properties of the material will determine the absolute temperature increase at the heating spot and the magnitude of the detected reflectivity at the probing spot. Thus, if light emerging from the heating laser or probe laser are polarized, this may affect the amplitude of the measured signal, but not the phase shift associated with the diffusive heat propagation in the sample.
 

\subsubsection*{Measuring Thermal Diffusivity}
The diffusive transport of heat is governed by the diffusion equation
\begin{equation}
\label{diffusioneq}
	\frac{\partial\,\delta T(\vec{r},t)}{\partial t}
-D\nabla^2\,\delta T(\vec{r},t)
=\frac{q(\vec{r},t)}{c}
\end{equation}
where $\delta T$ is the temperature disturbance above the ambient temperature $T$, $\vec{r}=(x_1, x_2, x_3)$ is the spherical radial coordinate given in terms of the euclidean principal axes $x_i$, $q$ is the absorbed power density, $c$ is the volumetric specific heat capacity, $D \equiv \kappa / c$ is the thermal diffusivity, and $\kappa$ is the thermal conductivity.  Note that $c$ and $D$ are themselves functions of $T$, but in the limit of weak heating $\delta T \ll T$, we make the approximation $c(T+\delta T)\approx c(T)$ and $D(T+\delta T)\approx D(T)$. Indeed, the temperature disturbance at the probing point from both lasers is estimated to be $\lesssim 1~$K through out the temperature range, so the above approximation is well justified, ensuring true linear response. Note that for the equation $D \equiv \kappa / c$ to hold, this requires that any modes in the specific heat $c$ not be zero modes. This is in addition to the two-phase model described in the main text.

Modulating the heating laser with a frequency $\omega$, we model the focused heat source as a point source, $q(t,\vec{r}\,)= P_0 e^{-i\omega t}\delta^3(\vec{r})$, which is valid as long as the distance from the heating spot is much larger than the spot radius.  The amplitude of the temperature variation at a point far from the heating spot is also expected to be modulated with the same frequency: $\delta T(\vec{r},t) = \tilde{\delta T}(\vec{r},\omega)e^{-i\omega t}$.  In a semi-infinite isotropic system, the temperature profile is spherically symmetric and takes the form
\begin{equation}
	\tilde{\delta T}(r,\omega)=\underbrace{\frac{P_0}{\kappa}
	\frac{1}{r}\exp \bigg(-\sqrt{\frac{\omega}{2D}}r\bigg)}_{\text{amplitude}}
		\underbrace{\exp\bigg(-i\sqrt{\frac{\omega}{2D}}r\bigg).}_{\text{phase}}
\label{diffsol}
\end{equation}

We vary the separation distance to verify the semi-infinite 3D system assumption and the small spot assumption, and we vary the heating power to verify the weak heating assumption. Our measurement gives us the response at the modulation frequency $\omega$. Because the separation distance between the lasers spots was measured using camera vision and each individually mounted sample comes at a small random tilt, a systematic error on the order of 5\% is associated to each set of measurements. This systematic error is constant and figures in the main text show this representative error on selected data points at the highest and lowest temperature of each measurement. 

Starting from Eqn.~\ref{diffsol}, we can write the instantaneous reflectivity at the probe point, which is held at an average temperature $T$ as:
\begin{equation}
\begin{split}
\delta R(r,t,T) &= \Re\left\{ A\frac{dR}{dT}\tilde{\delta  T}(r,\omega) e^{-i\omega t} \right\} \\
 &= A\frac{dR}{dT}\frac{P_0}{\kappa r}e^{- \phi(r,\omega)}\cos\left(\omega t + \phi(r,\omega)\right)
\end{split}
\label{delref}
\end{equation}
Here $A$ is an efficiency factor, which may depend  on mechanical vibrations, fluctuations in the laser power, and surface imperfections among other possible effects. At the same time  the phase shift of the signal:
\begin{equation}
\phi(r,\omega)= \sqrt{\frac{\omega}{2D}}r
\label{diphase}
\end{equation}
is the robust quantity to measure, which is independent of the local amplitudes either at the heating point at $r=0$ or the probe point at $r$, and only depend on the heat transport between these two points, making the direct measurement of the phase shift a robust way to extract the directional diffusivity. 

\begin{figure}[ht]
	\centering
	\includegraphics[width=01.0\columnwidth]{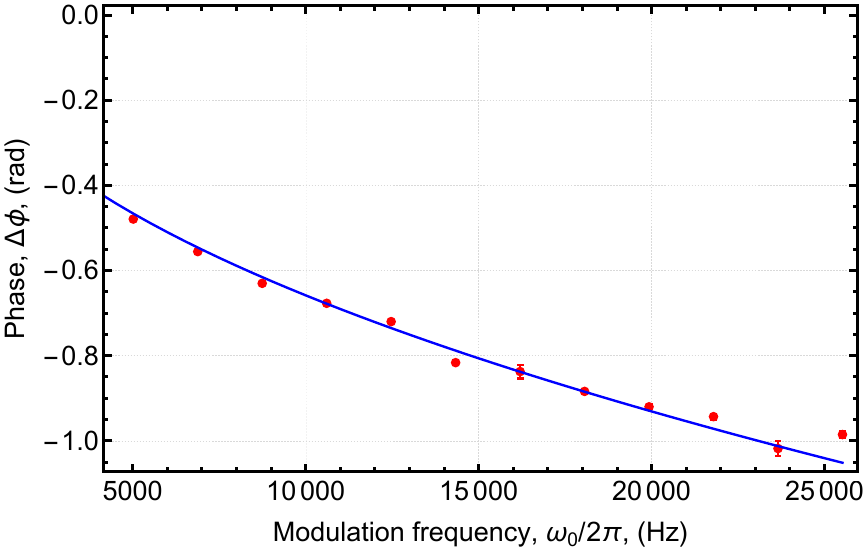}
	\caption{\label{phase}A typical phase delay as a function of heating laser modulation frequency, $\phi(\omega)$(red dot and error bars) obtained by sweeping $\omega$. This particular set is measured on CsV$_3$Sb$_5$ at a separation distance of $9.99~\mu$m at $110~$K. The blue line is a fit using the form in Eqn.~\ref{diphase}.}
\end{figure}

We obtain $D$ by fitting the phase delay $\phi$ between the source and the response signals as a function of $\omega$ at fixed $r$: $D = \omega r^2 / 2 \phi^2$. A typical fit is shown in Fig.~\ref{phase}. As Fig.~\ref{phase} shows and as repeated with multiple separation distances, the measured phase delays at a fixed separation distance as a function of heating laser modulation frequency follows the same function as the semi-infinite 3D system and small spot assumption (phase delay proportional to square root of modulation frequency) and so diffusivity can be extracted using  Eqn.~\ref{diffsol}. Because the amplitude of the reflected signal decreases at higher frequencies and because the amplitude of the reflected signal decreases at lower temperatures, measurements of diffusivity were taken by measuring the phase delay at 12 evenly spaced frequencies between $5.5~$kHz and $25.5~$kHz above 115~K and 3 evenly spaced frequencies between $5.5~$kHz and $8.7~$kHz below 115~K and fitting for diffusivity using Eqn.~\ref{diphase}. Multiple measurements of diffusivity at a given temperature were averaged to a single value.
We check the homogeneity of the crystals by repeating measurements at different positions on the surface, and check the isotropy/anisotropy by rotating the relative orientation between the laser spots. We also verify that there is no polarization dependence in the extracted diffusivity by repeating measurements at different heating laser and probing laser polarizations.



\end{document}